\pdfoutput=1
\documentclass[a4paper,american,floatfix,pdftex,superscriptaddress,twoside,%
aip,jcp,%
citeautoscript,
longbibliography,%
reprint,twocolumn,
]{revtex4-1}%
\usepackage{amsfonts,amsmath,amssymb}
\usepackage{commath}
\usepackage[T1]{fontenc}
\usepackage{lmodern}
\usepackage{graphicx}%
\usepackage[utf8]{inputenc}
\usepackage{microtype}
\usepackage[obeyFinal,textsize=footnotesize]{todonotes}
\usepackage{xspace}
\usepackage{hyperref, hypernat}
\usepackage[todos]{CMImacros}

\newcommand{\cfeldesy}{\affiliation{Center for Free-Electron Laser Science, Deutsches
      Elektronen-Synchrotron DESY, Notkestraße 85, 22607 Hamburg, Germany}}%
\newcommand{\uhhcui}{\affiliation{The Hamburg Center for Ultrafast Imaging, Universität Hamburg,
      Luruper Chaussee 149, 22761 Hamburg, Germany}}%
\newcommand{\uhhphys}{\affiliation{Department of Physics, Universität Hamburg, Luruper Chaussee 149,
      22761 Hamburg, Germany}}%
\newcommand{\ayemail}{\email[Email: ]{andrey.yachmenev@cfel.de}}%
\newcommand{\cmiweb}{\homepage[URL: ]{https://www.controlled-molecule-imaging.org}}%

\begin{document}
\title{Laser-induced dynamics of molecules with strong nuclear quadrupole coupling}%
\author{Andrey Yachmenev}\ayemail\cmiweb\cfeldesy\uhhcui%
\author{Linda V. Thesing}\cfeldesy\uhhcui\uhhphys%
\author{Jochen Küpper}\cfeldesy\uhhcui\uhhphys%
\date{\today}
\begin{abstract}
   We present a general variational approach for computing the laser-induced rovibrational dynamics
   of molecules taking into account the hyperfine effects of the nuclear quadrupole coupling. The
   method combines the general variational approach TROVE (Theoretical Ro-Vibrational Energies), 
   which provides accurate rovibrational
   hyperfine energies and wave functions for arbitrary molecules, with the variational method
   RichMol, designed for generalized simulations of the rovibrational dynamics in the presence of
   external electric fields. We investigate the effect of the nuclear quadrupole coupling on the
   short-pulse laser alignment of a prototypical molecule CFClBrI, which contains nuclei with large
   quadrupole constants. The influence of the nuclear quadrupole interactions on the post-pulse
   molecular dynamics is negligible at early times, first several revivals, however at longer
   timescales the effect is entirely detrimental and strongly depends on the laser intensity. This
   effect can be explained by dephasing in the laser-excited rotational wavepacket due to irregular
   spacings between the hyperfine-split nuclear spin states across different rotational hyperfine
   bands.
\end{abstract}
\maketitle

Laser-controlled rovibrational molecular dynamics is a subject of active research in physics and
chemistry~\cite{Christensen:PRL113:073005, Shepperson:PRL118:203203, Karamatskos:NatComm10:3364,
   Koch:RMP91:035005}. In particular, the control of molecular spatial alignment and
orientation~\cite{Stapelfeldt:RMP75:543, Holmegaard:PRL102:023001, Ghafur:NatPhys5:289} is highly
leveraged in many ultrafast imaging experiments~\cite{Hensley:PRL109:133202, Kuepper:PRL112:083002,
   Itatani:Nature432:867, Kanai:Nature435:470, Holmegaard:NatPhys6:428, Filsinger:PCCP13:2076,
   Weber:PRL111:263601, Karamatskos:NatComm10:3364} and stereochemistry
studies~\cite{Kuipers:Nature334:420, Rakitzis:Science303:1852, Aquilanti:PCCP7:291} to increase the
experimental resolution. The mechanism of the alignment and orientation is tied to the driven
rotational dynamics of molecules in the ground vibrational state and described by
two-photon~\cite{Felker:JPC90:724, RoscaPruna:PRL87:153902} or
three-photon~\cite{Spanner:PRL109:113001, Lin:NatComm9:5134} Raman excitation processes. The
adiabatic alignment is induced by slowly increasing laser electric field, which creates a
directional potential trapping molecules in pendular states. Adiabatic alignment combined with a dc
electric field, produces the mixed-field orientation of molecules~\cite{Holmegaard:PRL102:023001,
   Nielsen:PRL108:193001, Trippel:PRL114:103003}. In the impulsive alignment regime using short
laser pulses, the created rotational wavepacket evolves with repeated alignment/orientation and
antialignment revivals. In all of these schemes, the temporal evolution of the rotational wavepacket
and the resulting revival structure is characteristic to the rotational-energy-level structure of
the molecule~\cite{Seideman:PRL83:4971, RoscaPruna:JCP116:6579, Trippel:PRL114:103003}. Typically,
this temporal wavepacket evolution is accurately predicted using the rigid-rotor or semirigid rotor
Hamiltonian models~\cite{Stapelfeldt:RMP75:543, Seideman:AAMOP52:289, Thesing:JCP146:244304,
   Thesing:PRA98:053412}.

Many heavy atoms, such as, for example, bromine, iodine, or platinum, have large nuclear
quadrupoles. In molecules, these create large hyperfine splittings, comparable or even larger than
the rotational-energy spacings. These strong hyperfine interactions arise from the coupling between
the quadrupole moments of nuclei and the electric field gradients produced by the distribution of
nuclei and electrons through the molecule. In such heavy-atom containing molecules the presence of
the manifold of hyperfine-split nuclear-spin states makes the temporal evolution of the
laser-excited rotational wavepacket, and thus the alignment revival structure, more
complicated~\cite{Thomas:PRL120:163202}. Therefore, a detailed understanding of the effect of these
nuclear-quadrupole interactions is an important ingredient for the control over the alignment of
heavy-atom-containing molecules, with corresponding implications for molecular-frame imaging
experiments. Heavy atoms are commonly utilized as strong scattering and absorption centers in x-ray
imaging experiments~\cite{Kuepper:PRL112:083002, Erk:Science345:288, Nass:JSR22:225}. They are also
exploited as good leaving groups in Coulomb-explosion velocity-map imaging of molecular
dynamics~\cite{Nevo:PCCP11:9912, Takanashi:PCCP19:19707}.

Here, we present a general variational approach for computing the field-driven rovibrational
dynamics of molecules including nuclear-quadrupole interactions. The present approach extends our
previously reported variational method for computing the nuclear-quadrupole hyperfine spectra of
small molecules~\cite{Yachmenev:JCP147:141101}. We developed a generalized methodology for computing
the matrix representations of various electric-multipole-moment tensor operators in the basis of
hyperfine wave functions. These tensor operators are used as building blocks of the molecule-field
interaction potential in simulations of the field-driven rovibrational dynamics, as implemented in
the computational approach RichMol~\cite{Owens:JCP148:124102}. To our knowledge, this is the first
attempt to generalized simulations of this kind. We demonstrate the effect by calculating the
one-dimensional alignment dynamics of the asymmetric CF$^{35}$Cl$^{79}$Br$^{127}$I molecule by a
short laser pulse.

In brief, in RichMol the time-dependent wavepacket $\Psi(t)$ is built from a superposition of the
field-free rovibrational wave functions $\ket{J,m_J,l}$
\begin{equation}
  \label{eq:psit}
  \Psi(t) = \sum_{J,m_J,l} c_{J,m_J,l}(t) \ket{J,m_J,l},
\end{equation}
where $J$ and $m_J$ denote the quantum numbers of the total rotational angular momentum operator
$\mathbf{J}$ and its projection onto the laboratory-fixed $Z$ axis, respectively. $l$ represents a
set of additional rotational and vibrational quantum numbers. The time-dependent coefficients
$c_{J,m_J,l}(t)$ are determined by a numerical solution of the time-dependent Schrödinger equation
using the time-evolution operator method. The total Hamiltonian consists of the sum of the molecular
rovibrational Hamiltonian $H_\text{rv}$, with the eigenfunctions $\ket{J,m_J,l} $, and the
molecule-field interaction potential $V(t)$, expanded in terms of molecular electric multipole
moment operators:
\begin{equation}
  \label{eq:ht}
  H(t) = H_\text{rv} - \mu_A E_A(t) - \frac{1}{2}\alpha_{AB}E_A(t)E_B(t) + \ldots
\end{equation}
$A$ and $B$ are Cartesian indices denoting the $X$, $Y$, and $Z$ axes in the laboratory frame and
the summation over all Cartesian indices is implicitly assumed. $E_A(t)$ is the $A$ Cartesian
component of the electric field vector, and $\mu_A$ and $\alpha_{AB}$ are the electronic
contributions to the molecular-frame electric-dipole-moment vector and polarizability tensor,
respectively. The interaction terms of higher expansion order, such as the first and second
hyperpolarizability tensors, can also in principle be included into the sum in
\eqref{eq:ht}~\cite{Owens:JCP148:124102}.

The time-evolution operator for the time step $\Delta t=t-t'$ is computed using the split-operator
method as
\begin{multline}
   \mathcal{U}(t,t') = e^{-i\frac{\Delta t}{2\hbar} H_\text{rv}} \left( e^{i\frac{\Delta t}{\hbar} \mu_A E_A(\frac{t+t'}{2})} \right. \\
   \left. \cdot e^{i\frac{\Delta t}{2\hbar} \alpha_{AB}E_A(\frac{t+t'}{2})E_B(\frac{t+t'}{2})}\cdot\ldots\right)  e^{-i\frac{\Delta t}{2\hbar} H_\text{rv}}.
   \label{eq:utt}
\end{multline}
$H_\text{rv}$ is diagonal in the basis of its eigenfunctions $\ket{J,m_J,l}$ and the diagonal
elements are the molecular rovibrational energies. The exponentials of the Cartesian tensor
operators $\mu_A,\alpha_{AB},\ldots$ are evaluated using an iterative approximation based on the
Krylov-subspace methods. The computational performance of the iterative methods depends crucially on
how efficient the matrix-vector products can be computed between the operator exponential and the
subspace vectors. In the following, we will use $T_A^{(\Omega)}$ to denote any index-symmetric
Cartesian tensor operator of rank $\Omega$ with $A$ being a multi-index labeling Cartesian
components in the upper simplex of the tensor in the laboratory frame. For example, for the dipole
moment, $\Omega=1$ and $A=X$, $Y$, or $Z$, and for the polarizability tensor, $\Omega=2$ and $A=XX$,
$XY$, $XZ$, $YY$, $YZ$, or $ZZ$.

The computations of the matrix-vector products can be significantly sped up by expressing the matrix
representation of Cartesian tensor operators $T_A^{(\Omega)}$ in a contracted tensor form:
\begin{equation}
  \label{eq:jml_ta_jml}
  \langle J',m_J',l' |T_{A}^{(\Omega)}| J,m_J,l \rangle
  = \sum_{\omega=0}^{\Omega} \mathcal{M}_{A,\omega}^{(J',m_J',J,m_J)} \mathcal{K}_{\omega}^{(J',l',J,l)}.
\end{equation}
The sum runs over all irreducible representations $\omega$ of the tensor and the matrices
$\mathcal{M}_{A,\omega}^{(J',m_J',J,m_J)}$ and $\mathcal{K}_{\omega}^{(J',l',J,l)}$ decouple the
laboratory-frame projections $A$ and quantum numbers $m_J$ and $m_J'$ from the molecular-frame
rovibrational quantum numbers $l$ and $l'$.

The explicit expressions for the $\mathcal{M}_{A,\omega}$ and $\mathcal{K}_{\omega}$ matrices depend
on the form of the field-free wave functions $\ket{J,m_J,l}$. We use the general-molecule
variational approach TROVE (Theoretical Ro-Vibrational
Energies)~\cite{Yurchenko:JMS245:126, Yachmenev:JCP143:014105,
   Yurchenko:JCTC13:4368, Chubb:JCP149:014101} to compute the field-free energies and wave functions
$\ket{J,m_J,l}$, which are represented by linear combinations of products of vibrational wave
functions $\ket{v}$ and symmetric-top rotational functions $\ket{J,m_J,k}$
\begin{equation}
  \label{eq:jml}
  \ket{J,m_J,l} = \sum_{v,k} c_{v,k}^{(J,l)}\ket{v}\ket{J,m_J,k}.
\end{equation}
$v$ denotes the composite vibrational quantum number and $k$ denotes the quantum number of the
molecular-frame $z$-projection of the rotational angular momentum operator. Using the wave functions
from \eqref{eq:jml}, the expressions for $\mathcal{M}_{A,\omega}$ and $\mathcal{K}_{\omega}$ can be
derived as
\begin{multline}
   \mathcal{M}_{A,\omega}^{(J',m_J',J,m_J)} = (-1)^{m_J'}\sqrt{(2J'+1)(2J+1)} \\
   \times\sum_{\sigma=-\omega}^{\omega} [U^{(\Omega)}]^{-1}_{A,\omega,\sigma}
   \left(\begin{array}{ccc}J&\omega&J'\\m_J&\sigma&-m_J'\end{array}\right)
  \label{eq:m}
\end{multline}
and
\begin{multline}
  \label{eq:k}
  {\mathcal{K}_{\omega}^{(J',l',J,l)} = \sum_{k',v'}\sum_{k,v}
     \left[c_{v',k'}^{(J',l')}\right]^* c_{v,k}^{(J,l)}  (-1)^{k'}} \\
  {\times\sum_{\sigma=-\omega}^{\omega}\sum_{a}
     \left(\begin{array}{ccc}J&\omega&J'\\k&\sigma&-k'\end{array}\right)
     U^{(\Omega)}_{\omega,\sigma,a}\braopket{v'}{T_{a}^{(\Omega)}}{v}}.
\end{multline}
$T_{a}^{(\Omega)}$ denotes Cartesian tensor operator in the molecular frame, with $a$ being a
Cartesian multi-index (similar to $A$), and the matrix $U^{(\Omega)}$ defines the transformation of
tensor from Cartesian to spherical-tensor form~\cite{Owens:JCP148:124102}.

This approach is general and can be interfaced with any rovibrational computer code that provides
the tensor-matrix elements in the form of \eqref{eq:jml_ta_jml}. The approach also permits the use
of more complex field-free wave functions than those defined in \eqref{eq:jml}. As multipole-moment
operators commute with the nuclear-spin angular momenta, the operators' matrix elements in the basis
of hyperfine states can also be cast into the form of \eqref{eq:jml_ta_jml}. In the following, we
derive the explicit expressions for the $\mathcal{M}_{A,\omega}$ and $\mathcal{K}_{\omega}$ matrix
elements in the basis of the nuclear-spin hyperfine-structure wave functions.

The general variational implementation of the nuclear-spin hyperfine effects at the level of the
nuclear-quadrupole interaction was recently implemented~\cite{Yachmenev:JCP147:141101} and used for
the generation of a quadrupole-resolved line list of the ammonia molecule~\cite{Coles:ApJ870:24}.
The nuclear-quadrupole interaction in a molecule containing $n=1\ldots{}N$ quadrupolar nuclei is
described by the coupling of the electric field gradient tensor (EFG) at each $n$-th nucleus
$\mathbf{V}(n)$ with its quadrupole moment tensor $\mathbf{Q}(n)$. The total spin-rovibrational
Hamiltonian takes the form
\begin{align}
  \label{eq:hsrv}
  H_\text{srv} = H_\text{rv} + \sum_{n} {\bf V}(n)\cdot\mathbf{Q}(n).
\end{align}
The overall rotational, $\mathbf{J}$, and the nuclear spin, $\mathbf{I}_n$, and total, $\mathbf{F}$,
angular momentum operators for $n=1\ldots{}N$ quadrupolar nuclei are coupled as
$\mathbf{I}_{1,2}=\mathbf{I}_{1}+\mathbf{I}_{2}$,
$\mathbf{I}_{1,3}=\mathbf{I}_{1,2}+\mathbf{I}_{3}$, \ldots,
$\mathbf{I}_{1,N-1}=\mathbf{I}_{1,N-2}+\mathbf{I}_{N-1}$,
$\mathbf{I}\equiv\mathbf{I}_{1,N}=\mathbf{I}_{1,N-1}+\mathbf{I}_{N}$, and
$\mathbf{F}=\mathbf{J}+\mathbf{I}$. The nuclear-spin functions $\ket{I,m_I,\mathcal{I}}$ depend on
the quantum numbers $I$ and $m_I$ of the collective nuclear spin angular momentum operator
$\mathbf{I}$ and its projection onto the laboratory $Z$ axis, respectively. The set of auxiliary
spin quantum numbers $\mathcal{I}=\{I_{1},I_{1,2},\ldots,I_{1,N-1}\}$ of the intermediate spin
angular momentum operators provide a unique assignment of each nuclear-spin state. The total
spin-rovibrational wave functions $\ket{F,m_F,u}$ are built as symmetry-adapted linear combinations
of the products of the rovibrational wave functions $\ket{J,m_J,l}$ in \eqref{eq:jml} and the
nuclear-spin functions $\ket{I,m_I,{\mathcal I}}$:
\begin{multline}
   \ket{F,m_F,u} = \sum_{I,\mathcal{I},J,l} c_{I,\mathcal{I},J,l}^{(F,u)} \left[ \sum_{m_J,m_I} (-1)^{F+m_F}\right. \\
   \times\left. \sqrt{2F+1} \left(\begin{array}{ccc} F & I & J \\ -m_F & m_I & m_J \\ \end{array} \right)
      \ket{I,m_I,\mathcal{I}} \ket{J,m_J,l} \right].
   \label{eq:fmfu}
\end{multline}
$c_{I,\mathcal{I},J,l}^{(F,u)}$ are the eigenvector coefficients of the spin-rovibrational
Hamiltonian \eqref{eq:hsrv}, $F$ and $m_F$ are the quantum numbers of $\mathbf{F}$ and its
projection onto the laboratory-fixed $Z$-axis, and $u$ denotes the hyperfine-state running index.

The explicit expressions for the matrix elements of the spin-rovibrational Hamiltonian
\eqref{eq:hsrv} for an arbitrary number of quadrupolar nuclei and details of the variational
solution can be found in reference~\onlinecite{Yachmenev:JCP147:141101}. Here, we derived explicit
expressions for the matrix elements of the general electric multipole Cartesian tensor operator
$\braopket{F',m_F',u'}{T_A^{(\Omega)}}{F,m_F,u}$, which in the contracted-tensor form of
\eqref{eq:jml_ta_jml} are
\begin{multline}
   \mathcal{M}_{A,\omega}^{(F',m_F',F,m_F)} = (-1)^{m_F'}\sqrt{(2F'+1)(2F+1)} \\
   \times\sum_{\sigma=-\omega}^{\omega} [U^{(\Omega)}]^{-1}_{A,\omega,\sigma}
   \left(\begin{array}{ccc}F'&\omega&F\\-m_F'&\sigma&m_F\end{array}\right)
  \label{eq:mf}
\end{multline}
and
\begin{multline}
   {\mathcal{K}_{\omega}^{(F',u',F,u)}
      = \sum_{I',{\mathcal I}',J',l'}\sum_{I,{\mathcal I},J,l} \left[ c_{I',{\mathcal I}',J',l'}^{(F',u')} \right]^* c_{I,\mathcal{I},J,l}^{(F,u)} (-1)^{I}} \\
   {\times\sqrt{(2J'+1)(2J+1)}\left\{\begin{array}{ccc}J'&F'&I \\ F&J&\omega\end{array}\right\}
      \mathcal{K}_{\omega}^{(J',l',J,l)}\delta_{I',I}\delta_{\mathcal{I}',\mathcal{I}}}.
   \label{eq:kf}
\end{multline}
$\mathcal{K}_{\omega}^{(J',l',J,l)}$, defined in \eqref{eq:k}, contains the rovibrational matrix
elements of $T_A^{(\Omega)}$ in the basis $\ket{J,m_J,l}$. Using the expressions \eqref{eq:mf} and
\eqref{eq:kf} in \eqref{eq:jml_ta_jml}, RichMol could directly be employed to simulate the coupled
nuclear-spin-rovibrational molecular dynamics in external fields. In \eqref{eq:utt}, the diagonal
representation of $H_\text{rv}$ in the rovibrational energies was replaced by the diagonal
representation of $H_\text{srv}$ in the hyperfine energies \eqref{eq:hsrv}.

This new approach was used to investigate the effect of nuclear-quadrupole coupling on the impulsive
alignment of bromochlorofluoroiodomethane CF$^{35}$Cl$^{79}$Br$^{127}$I. This molecule has a
quasi-rigid structure and contains three different heavy nuclei with large quadrupole coupling
constants and with correspondingly nontrivial laser-induced rotational and nuclear spin spectra and
dynamics. A short non-resonant laser pulse is linearly polarized along the laboratory $Z$ axis. Its
intensity is given by the Gaussian function $I(t)=I_0\exp\left(-4\ln2t^2/\sigma^2\right)$ with
$\sigma=1~\text{ps}$ and $I_0=6\times 10^{11}~\Wpcmcm$ or $1\times10^{12}~\Wpcmcm$. The excitation
by a non-resonant laser field is described by the electric polarizability term in the interaction
potential \eqref{eq:ht}. The explicit expressions for the elements of the matrix $U^{(2)}$ for the
polarizability are listed in the Table~I of reference~\onlinecite{Owens:JCP148:124102}. The degree
of molecular alignment is characterized by the expectation value
$\costhreeD=\braopket{\Psi(t)}{\cos^2\theta}{\Psi(t)}$, with the angle $\theta$ between the
molecular-frame $z$ and the laboratory-frame $Z$ axes. The matrix elements of the $\cos^2\theta$
operator can be easily computed using the general expressions \eqref{eq:mf}, \eqref{eq:kf}, and
\eqref{eq:k} by noting the relationship $\cos^2\theta = (2d_{00}^{2}+1)/3$, where $d$ is the
Wigner~$d$ matrix. The vibrational matrix elements of $d_{00}^{2}$ in \eqref{eq:k} are
$\langle v' |d_{00}^{2}|v\rangle=\delta_{v'v}$ and
$U_{\omega,\sigma}^{(0)}=\delta_{\omega,2}\delta_{\sigma,0}$; here the index $a$ in \eqref{eq:k} is
redundant.

For simplicity we neglected the vibrational motion of the molecule and approximated the full
rovibrational wave functions in \eqref{eq:jml} by rigid-rotor solutions. The rigid-rotor
approximation in simulations of the laser-induced alignment of quasi-rigid and even non-rigid
molecules in ultracold molecular beams has been validated in numerous
studies~\cite{Hamilton:PRA72:043402, Rouzee:PRA73:033418, Rouzee:PRA77:043412,
   Trippel:PRA89:051401R, Trippel:PRL114:103003, Thesing:PRA98:053412}. The zero-point vibrational
corrections to various electromagnetic tensors are known to be quite small, on the order of 1--3\%,
even for non-rigid molecules like H$_2$O$_2$~\cite{TorrentSucarrat:JCP122:204108} and have been
neglected in this study. The equilibrium geometry and polarizability of CFClBrI were calculated
using density functional theory (DFT) with the B3LYP functional and the def2-TZVPP basis
set~\cite{Weigend:JCP119:12753, Weigend:PCCP7:3297} in conjunction with the relativistic effective
core potential def2-ECP on the iodine atom~\cite{Peterson:JCP119:11113}. The accuracy of hybrid
functionals for the prediction of molecular polarizabilities was assessed on a dataset of
132~molecules, yielding a root-mean-square error of 3--5~\% relative to coupled-cluster
singles and doubles with a perturbative correction to triples
[CCSD(T)]~\cite{Hait:PCCP20:19800}. Calculations of the electric field gradient tensors, needed for
the nuclear-quadrupole coupling Hamiltonian in \eqref{eq:hsrv}, were carried out at the DFT/B3LYP
level of theory using the all-electron scalar relativistic Douglas-Kroll-Hess
Hamiltonian~\cite{Neese:JCP122:204107} with the DKH-def2-TZVP basis set~\cite{Jorge:JCP130:064108,
   Campos:MolPhys111:167}. Systematic studies of the accuracy of DFT functionals for predictions of
the electric field gradient tensors of transition metal complexes provided a mean error estimate of
about 0.071~a.u. for the B3LYP functional~\cite{Bjornsson:CPL559:112}. All electronic structure
calculations employed the quantum chemistry package ORCA~\cite{Neese:WIRCMS2:73,
   Neese:WIRCMS8:e1327}. The quadrupole moments for the $^{35}$Cl, $^{79}$Br, and $^{127}$I nuclei
are $Q=-81.65$~mb, $313$~mb, and $-696$~mb, respectively~\cite{Pyykko:MolPhys106:1965}.

\begin{figure*}
   \includegraphics[width=\linewidth]{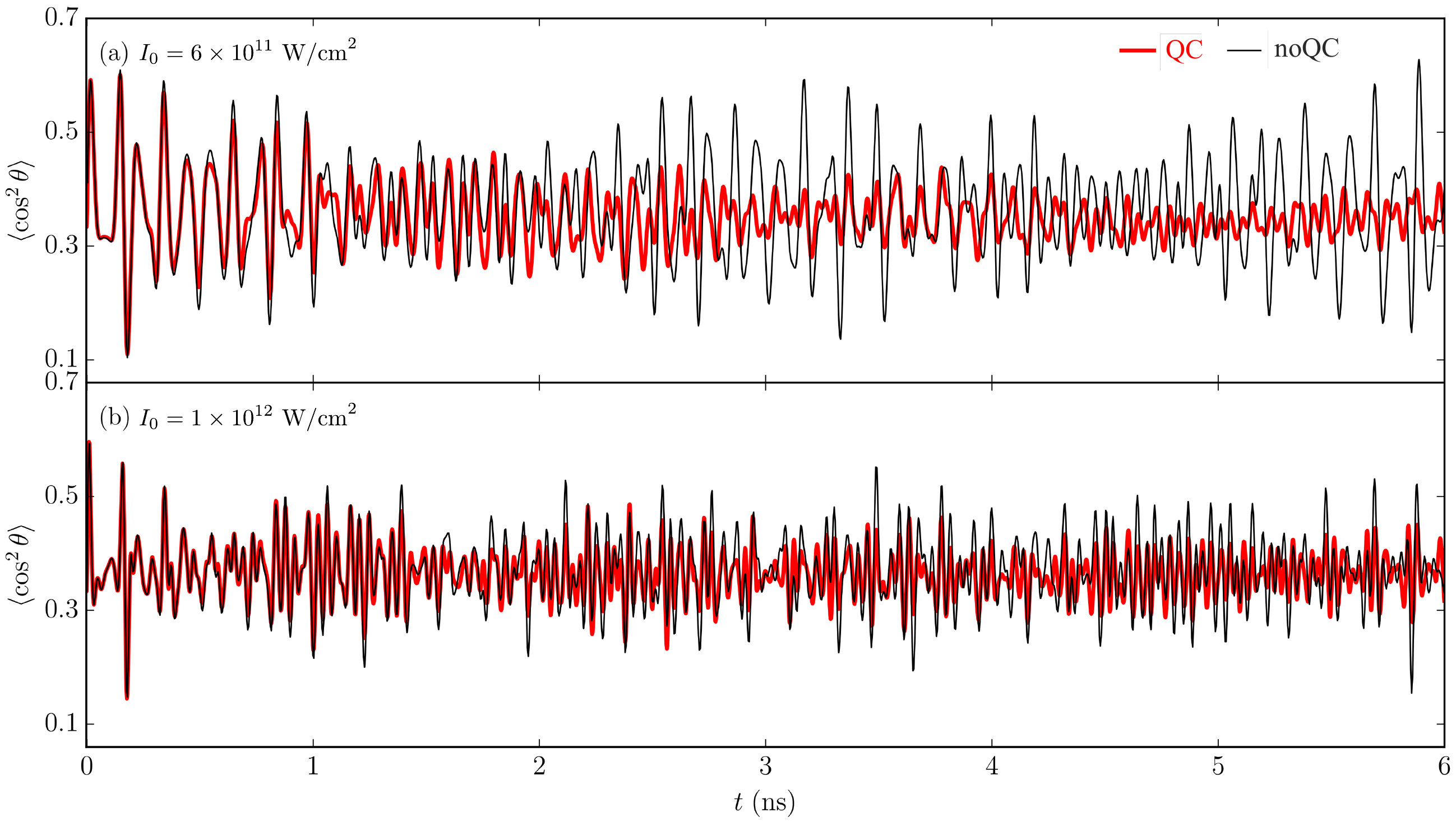}
   \caption{Laser-alignment dynamics of CFClBrI, with and without nuclear-quadrupole coupling (QC
      and noQC) following excitation by a Gaussian laser pulse with $\tau_\text{FWHM}=1~\text{ps}$
      and a maximum laser intensity of (a) $I_0=6\times10^{11}~\Wpcmcm$ and (b)
      $I_0=1\times 10^{12}~\Wpcmcm$.}
   \label{fig:results}
\end{figure*}

The calculations of the molecule-field dynamics in the basis of the nuclear-spin hyperfine states
were performed in a three-step procedure: First we solved the field-free problem using the
rigid-rotor Hamiltonian and obtained the rotational energies and the rotational matrix elements of
the EFG and the polarizability tensors, as well as of the $\cos^2\theta$ operator, \eqref{eq:m} and
\eqref{eq:k}. Then we used the rotational energies and the matrix elements of the EFG tensors
together with the nuclear-quadrupole constants to build and diagonalize the nuclear-spin
quadrupole-coupling Hamiltonian~\cite{Yachmenev:JCP147:141101}, \eqref{eq:hsrv}. We obtained the
spin-rotational eigenfunctions in \eqref{eq:fmfu} and transformed the rotational matrix elements of
the polarizability tensor and $\cos^2\theta$ operator into the spin-rotational eigen-basis using
\eqref{eq:mf} and \eqref{eq:kf}. Finally, we used the spin-rotational energies and the matrix
elements of the polarizability to build the molecule-field interaction Hamiltonian and solve the
time-dependent problem. The wavepacket $\Psi(t)$ in \eqref{eq:psit} was built from a linear
combination of spin-rotational wave functions with time-dependent coefficients. The expectation
value \costhreeD is calculated from the spin-rotational matrix elements of $\cos^2\theta$ computed
at the previous step.

Here, we used a rotational basis that included all spin-rotational states of CFClBrI with
$F\leq41/2$, corresponding to $J\leq26$. We assumed that CFClBrI is initially in the ground
rotational state $J=0$, which has $96$ different nuclear-spin state components, including the $m_F$
degeneracy. Therefore, we performed a series of simulations starting from the different nuclear-spin
components and averaged the results assuming equal normalized populations of different spin
components. For comparison, we also performed calculations neglecting the hyperfine effects, \ie,
using rigid-rotor wavefunctions as the field-free basis. It should be noted that samples of heavy
molecules populating a few of the lowest rotational states, with $T_\text{rot}\approx0.4$~K, can be
produced from cold molecular beams using the electrostatic deflector~\cite{Filsinger:JCP131:064309,
   Chang:IRPC34:557, Trippel:RSI89:096110}.

The temporal evolution of the alignment calculated for two different laser intensities is shown
in~\autoref{fig:results}. For both laser intensities, the alignment without nuclear-quadrupole
interaction shows complex revival patterns originating from the dephasing and rephasing of the
rotational wavepacket. Since CFClBrI is an asymmetric-top molecule (asymmetry parameter
$\kappa=-0.75$), the revival patterns lack the typical periodicity of the $J$-type or $C$-type
revivals observed for near-symmetric-top molecules~\cite{Poulsen:JCP121:783, Rouzee:PRA73:033418,
   Holmegaard:PRA75:051403R}. Noticeably, the broader wave packet, produced by the higher-intensity
pulse, shows higher-frequency oscillations and generally reduced peak alignment. This is ascribed to
mismatches of the phases of the populated rotational states after the laser
pulse~\cite{Chatterley:JCP148:221105} and successive dephasing due to the molecule's
asymmetry~\cite{Rouzee:PRA73:033418}, preventing the simultaneous rephasing of more than a few
rotational states with significant populations.

The nuclear-quadrupole interaction increases complexity of the revival dynamics and depletion of the
peak alignment compared to the rigid-rotor results. Nevertheless, during the first 300~ps after the
laser pulse the degree of alignment is almost identical to the rigid-rotor result. We also observed
a dependence of the nuclear-quadrupole effect on the laser-field intensity. For the lower intensity,
\autoref[(a)]{fig:results}, the impact is seen about 200~ps earlier than for the higher intensity,
\autoref[(b)]{fig:results}.

The non-periodic recurrences in the alignment dynamics of the asymmetric-top molecules originate
from large asymmetry splittings of the rotational energy levels. The influence of the
nuclear-quadrupole interaction can be understood in a similar manner, as dephasing effect resulting
from the incommensurate hyperfine splittings for different rotational states. The dephasing is
stronger for the wavepacket dominated by the low-energy rotational states, where the hyperfine
splittings are most irregular. For large $J$, the hyperfine splittings become increasingly
uniform~\cite{Gordy:MWMolSpec} and the contribution to the dephasing is minimized. This explains why
the effect of the nuclear-quadrupole interaction is less striking for the higher-intensity laser
field, which populates higher-energy rotational states. For both intensities, however, the small-$J$
states have relatively large populations and the nuclear-quadrupole interaction has a strong
influence on the field-free alignment.

In conclusion, we have presented the first general implementation of nuclear-quadrupole hyperfine
effects in the rovibrational dynamics of molecules driven by external electric fields. Our approach
combines TROVE, which provides hyperfine energies and wave functions for arbitrary molecule, with
RichMol, designed for generalized simulations of external-field effects. In principle, the presented
approach can be applied to simulate the fully-coupled spin-rovibrational dynamics of any molecule
with no inherent limitations on the number of quadrupolar nuclei. The external field effects are not
limited by the polarizability interaction: other multipole moment operators, including the permanent
dipole moment or the first and second hyperpolarizabilities, can be considered without additional
implementation efforts.

We studied the influence of the nuclear-quadrupole coupling on the laser impulsive alignment of a
prototypical heavy-atom molecule CFClBrI. While the effect is small for the first few revivals, it
turns out to be entirely detrimental for the revivals at later times. The laser field plays an
important role, with lower intensities prompting larger effect of the nuclear-quadrupole coupling.
This can be explained by dephasing of the rotational wavepacket due to the incommensurate structure
of the hyperfine-split levels for different rotational states. This effect is stronger for small-$J$
rotational states and practically disappears for states with high angular momenta. Given that
small-$J$ states in the wavepacket will be the source of the largest dephasing effects, we expect
that stronger laser intensities and higher initial rotational temperatures should further diminish
the effect of the nuclear-quadrupole coupling. More rigorous studies of these effects for different
alignment scenarios and molecular systems are ongoing. We envisage future applications of the
presented approach to inform and interpret diverse laser-field experiments on molecules containing
nuclei with large quadrupole constant.

This work has been supported by the Deutsche Forschungsgemeinschaft (DFG) through the priority
program ``Quantum Dynamics in Tailored Intense Fields'' (QUTIF, SPP1840, KU~1527/3, YA~610/1) and
the clusters of excellence ``Center for Ultrafast Imaging'' (CUI, EXC~1074, ID~194651731) and
``Advanced Imaging of Matter'' (AIM, EXC~2056, ID~390715994). L.T.\ thanks Rosario
Gonz{\'a}lez-F{\'e}rez for helpful discussions and for the hospitality at the University of Granada.

\bibliography{string,cmi}
\end{document}